\documentclass[11pt,tightenlines,notitlepage]{revtex4-1}
\usepackage{amsmath}    
\usepackage{graphicx}   
\usepackage{epstopdf}

\newcommand{\vko}{\,\vec{k}_{o}}
\newcommand{\ko}{\, k_{o}}
\newcommand{\kox}{\,k_{o,x}}
\newcommand{\koz}{\,k_{o,z}}
\newcommand{\vkt}{\,\vec{k}_{t}}
\newcommand{\kt}{\, k_{t}}
\newcommand{\ktx}{\,k_{t,x}}
\newcommand{\ktz}{\,k_{t,z}}
\newcommand{\xh}{\,\hat{x}}
\newcommand{\zh}{\,\hat{z}}
\newcommand{\nm}{\, \rm{nm}}
\newcommand{\Phiz}{\, \Phi_{z}}
\newcommand{\PhizTM}{\, \Phi_{z,TM}}
\newcommand{\PhizTE}{\, \Phi_{z,TE}}

\newcommand{\zo}{\, z_{0}}

\newcommand{\di}{\, d_{i}}
\newcommand{\e}{\, \epsilon}
\newcommand{\ce}{\, \underline{\epsilon}}

\newcommand{\ei}{\, \epsilon_{i}}
\newcommand{\eM}{\, \epsilon_{M}}
\newcommand{\ceM}{\, \underline{\epsilon}_{M}}
\newcommand{\eD}{\, \epsilon_{D}}
\newcommand{\dM}{\, d_{M}}
\newcommand{\dD}{\, d_{D}}
\newcommand{\cei}{\, \underline{\epsilon}_{i}}

\newcommand{\cn}{\, \underline{n}}

\newcommand{\ckz}{\, \underline{k}_{z}}
\newcommand{\ct}{\, \underline{t}}
\newcommand{\cp}{\, \underline{p}}
\newcommand{\cgM}{\, \underline{\gamma}_{M}}
\newcommand{\lo}{\, \lambda_{0}}

\graphicspath{ {C:/Users/ott/Documents/MATLAB/UBC/TMM_oop/} }

\begin{document}

\title{Flat Lens Criterion by Small-Angle Phase}
\author{Peter Ott,$^1$ Mohammed H. Al Shakhs,$^2$ Henri J. Lezec,$^3$ and Kenneth J. Chau$^2$}
\affiliation{$^1$Heilbronn University, Heilbronn, Germany\\
$^2$School of Engineering, The University of British Columbia, Kelowna, British Columbia, Canada\\
$^3$Center for Nanoscale Science and Technology, National Institute of Standards and Technology, Gaithersburg, Maryland, USA}


\begin{abstract}
We show that a classical imaging criterion based on angular dependence of small-angle phase can be applied to any system composed of planar, uniform media to determine if it is a flat lens capable of forming a real paraxial image and to estimate the image location.  The real paraxial image location obtained by this method shows agreement with past demonstrations of far-field flat-lens imaging and can even predict the location of super-resolved images in the near-field. The generality of this criterion leads to several new predictions: flat lenses for transverse-electric polarization using dielectric layers, a broadband flat lens working across the ultraviolet-visible spectrum, and a flat lens configuration with an image plane located up to several wavelengths from the exit surface.  These predictions are supported by full-wave simulations.  Our work shows that small-angle phase can be used as a generic metric to categorize and design flat lenses.
\end{abstract}

\maketitle

\section{Introduction}

Glass lenses found in cameras and eyeglasses have imaging capabilities derived from the shapes of their entrance and exit faces.  Under certain conditions, it is possible to image with unity magnification using a perfectly flat lens constructed from planar, homogeneous, and isotropic media.  Unlike other lenses that are physically flat (such as graded-index lenses or meta-screens), a flat lens has complete planar symmetry and no principle optical axis, which affords the unique possibility of imaging with an infinite aperture.

Flat lenses for far-field and near-field operation have been proposed by respective consideration of propagating and evanescent waves (Fig. 1).  Optical ray visualization of propagating waves suggests that a far-field flat lens requires a planar homogeneous medium that is either isotropic and negative index (known as a Veselago lens named after its originator~\cite{Veselago1968}) or anisotropic with a constitutive tensor having diagonal components of opposite sign~\cite{Smith2004,Lu2005,Dumelow2005}.  A solution to Maxwell's equations indicates that a planar, negative-index medium is capable of amplifying evanescent waves, which could enable imaging with unlimited resolution~\cite{Pendry2000}.  In the absence of naturally-occurring negative-index materials, it was suggested that a thin layer of silver could also amplify evanescent waves, although restricted to transverse-magnetic (TM) polarization and confined to the near field~\cite{Pendry2000}.

In recent years, flat lenses for TM polarization have been demonstrated using a single metallic layer~\cite{Pendry2000,Blaikie2002,Platzman2002,Zhang2003,Melville2004,Melville2004B,Durant2005,Zhang2005Science,Melville2005,Zhang2005NJP} and metal-dielectric multi-layers~\cite{Solymar2001,Pendry2003,Belov2006,Webb2006,Elson2006,Melville2007,Scalora2008,Moore2008,Moore2009,Kotynski2009,Kotynski2011JAP,Kotynski2011APA,Benedicto2012,Lezec2013}.  There have been a variety of methods used to design these systems.  Some implementations are based on mimicking physical processes (evanescent wave amplification~\cite{Melville2004,Melville2004B,Zhang2005Science,Melville2005,Zhang2005NJP}) or conditions (such as anisotropy~\cite{Pendry2003,Belov2006,Webb2006,Kotynski2011JAP} or negative index~\cite{Lezec2013})  described in past flat lens proposals.  Others use simulations~\cite{Scalora2008,Kotynski2009,Kotynski2011APA} or transfer function calculations~\cite{Elson2006,Melville2007,Moore2008,Moore2009}.  Recent work has also shown the possibility of designing flat lenses using the concept of band diagrams~\cite{Chau2014}.  As a complement to these studies, we introduce a flat lens criterion based on small-angle phase behavior that can be applied to any system composed of planar, homogeneous, and isotropic media to indicate the possibility of imaging and to estimate the image plane location. The criterion accurately predicts the image plane location for the far-field flat lens studied in~\cite{Lezec2013} and, rather surprisingly, yields image plane locations that are consistent with near-field flat-lens imaging presented in~\cite{Melville2004,Zhang2005Science,Melville2005,Belov2006,Kotynski2011APA}.  We derive approximate analytical expressions that delineate when a single layer and multi-layers act as flat lenses. Finally, small-angle phase is used to predict necessary conditions to realize flat lenses for TE polarization, a broadband flat lens operating over the ultraviolet-visible spectrum and an immersion flat lens with an adjustable image plane location.  Predictions are validated by full-wave numerical simulations, through which we discover that our proposed flat lens for TE polarization is also capable of super-resolution imaging.

\begin{figure}[h]
  \begin{center}
    \includegraphics[scale = 0.75]{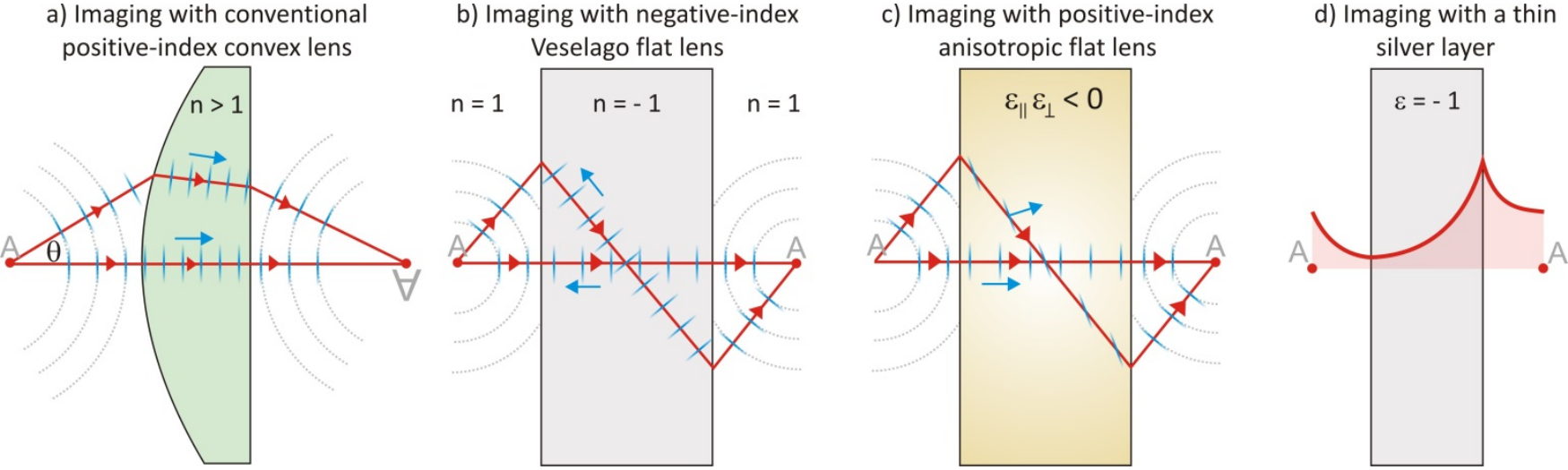}
    \caption[]{Optical ray visualization of imaging in (a) a standard plano-convex lens, (b) a planar negative-index slab, and (c) a planar anisotropic slab.  The red lines and blue arrows respectively indicate the local power and phase flow.  (d) Imaging in a thin silver layer by evanescent wave amplification. \label{Fig1}   }
  \end{center}
\end{figure}

\section{Flat Lens Criterion}

Classical paraxial image theory can be adapted to describe the imaging response of a flat lens to a point source using the concept of a pupil function (PF), where the size of the pupil is now permitted to be infinite.  As a supplement to a standard approach of flat lens analysis based on the PF amplitude~\cite{Solymar2001,Zhang2003,Pendry2003,Melville2007}, we emphasize the importance of PF phase as a potential indicator of a flat lens.  We begin by considering a planar medium of thickness $d$ with one face aligned to the $z=-d$ plane and translational symmetry in the $xy$ plane.  The object region ($z<-d$) and image region ($z > 0$) are composed of isotropic, homogeneous media that are generally dissimilar.  We consider a point source located in the object region directly on the entrance face of the medium at $z = -d$.  The source can be decomposed in the $xz$ plane as a uniform plane-wave spectrum parameterized by the wave vector in the image region $\vkt = \ktx\xh + \ktz\zh$.  Due to translational symmetry, we need only to evaluate the phase of the plane-wave components along the $z$ axis, $\Phiz$.  The phase of the plane-wave components exiting the slab, referenced to a common initial value at the source, define the PF phase $\Phiz(z=0)$ and map out loci of equi-phase points tracing out the wavefront in the image region.

The PF phase can be extrapolated to an arbitrary point $\zo$ by $\Phiz(\zo) = \Phiz(z=0) + \ktz \zo$.  Based on Fermat's principle, an ideal image forms at $\zo$ if the phase at $\zo$ is invariant for all plane-wave components, $\partial \Phiz(\zo)/\partial \ktz = 0$.  In the non-ideal case in which phase invariance for all plane-wave components is not achieved at any point, image formation suffers from spherical abberations.  By imposing phase invariance for just the small-angle plane-wave components ($\ktx << \ktz \simeq \kt$), we can define a paraxial image location in terms of the PF phase given by
\begin{equation}\label{paraxialimage}
s = \zo \equiv - \frac{\partial \Phiz(z=0)}{\partial \ktz}\Big\vert_{\ktz = \kt} \quad,
\end{equation}
which is positive for a real image in the image region and negative for a virtual image in the slab or object regions.  A flat lens can now be generally defined as a planar medium capable of producing a real paraxial image of a point source located on its entrance face, a condition met when $s>0$.  This flat lens criterion, which  is defined for numerical aperture (NA) near zero, serves as an analogue to paraxial image theory used to describe conventional optical systems.  Because this condition depends only on the phase profile at the slab exit, it can be used as a generic indicator of imaging applicable to any flat lens.

PF amplitude is commonly displayed versus the normalized lateral wave vector $\ktx/\kt$, which is consistent with interpretations of the PF as a transfer function relating field quantities Fourier transformed along iso-planatic object and image planes perpendicular to the optical axis~\cite{Goodman1968}.  PF phase, on the other hand, is not conveniently displayed versus $\ktx/\kt$ because spherical wavefront curvature and spherical abberations respectively map onto quadratic and higher even-order terms that are difficult to distinguish on a graph.  Given the dependence of $\Phiz(z=0)$ on $\ktz$ in Eq.~(\ref{paraxialimage}), we propose to display the PF phase as a function of $q_t= 1 - \ktz/\kt = 1 - \cos \theta_t$ (where $\theta_t$ is the angle in the image region).  Similar to $\ktx/\kt$, $q_t$ describes propagating waves over the range $0 \leq q_t \leq 1$ (Fig. 2).  By plotting PF phase versus $q_t$, a positive slope reveals flat lens behavior, the paraxial image location can be estimated by the slope at the $y$-intercept ($\partial \Phiz(z=0)/\partial q_t \vert_{q_t=0} $), and spherical abberations can be identified by departures from linearity.

We have considered an ideal object consisting of a point source located on the entrance face of the slab.  The analysis can be extended to volumetric objects treated as collections of point sources, whose image locations are found by applying the flat lens criterion to a point source scanned in the lateral and longitudinal directions.   Due to translational symmetry of the slab, a point source can be translated in the lateral direction with impunity.  Longitudinal translation of a point source away from the slab entrance can be described by adding an additional layer between the source and the entrance.  Shifts of the point source along the lateral or longitudinal directions map onto similar shifts of the image location.  For this reason, the image location $s$ also approximates the maximum working distance of the flat lens.  That is, a flat lens can generate a real image of an object pressed against its entrance so long as the depth of the object is less than $s$.

\begin{figure}[h]
  \begin{center}
    \includegraphics[scale = 0.58]{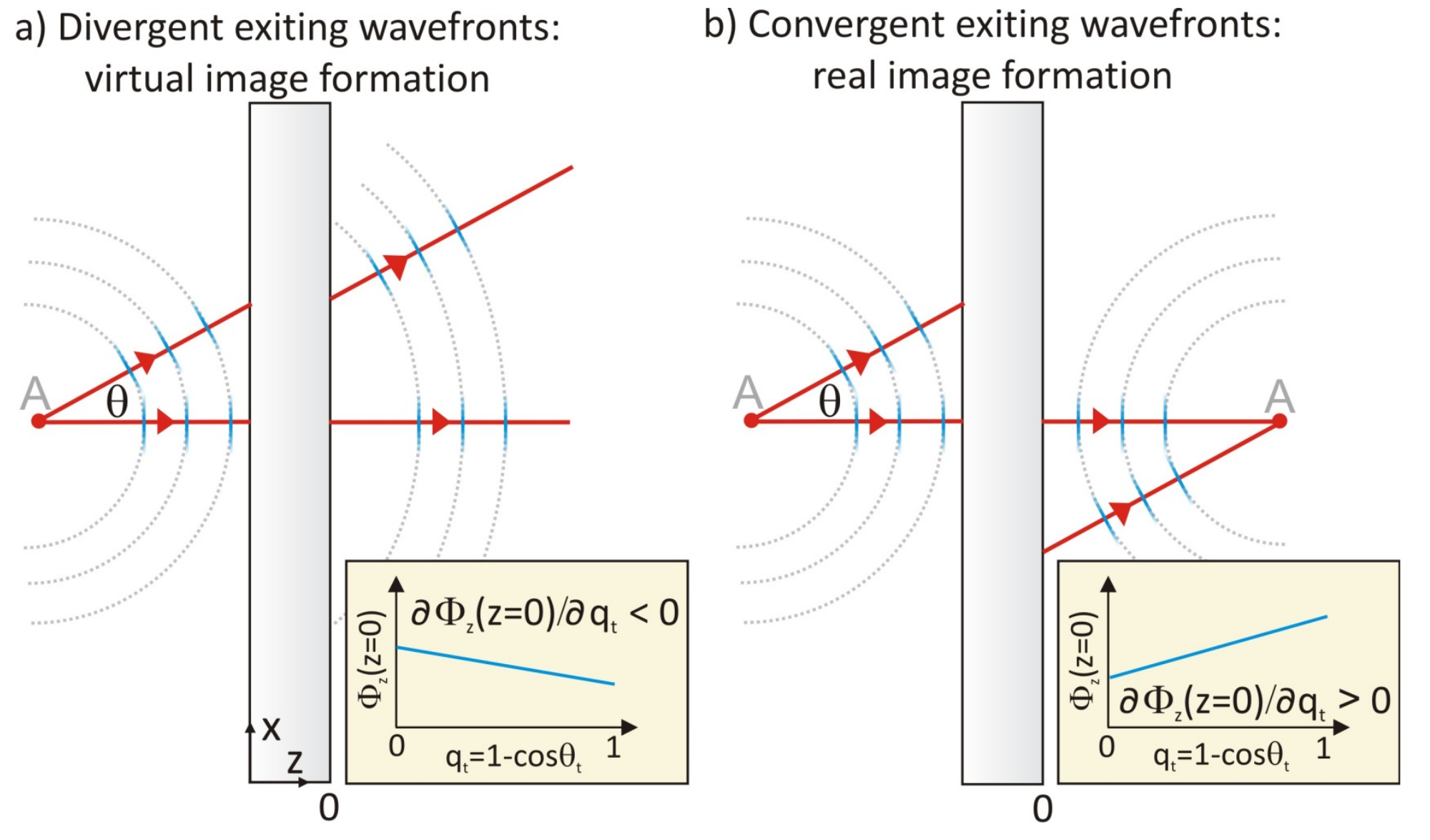}
    \caption[]{Curvature of the wavefronts exiting a planar slab for the cases of (a) virtual and (b) real image formation.  By plotting the phase $\Phiz(z=0)$ versus $q_t = 1 - \cos \theta_t$, the possibility of a flat lens can be determined by inspection from a positive slope. \label{Fig2}}
  \end{center}
\end{figure}

\section{Methodology}

We first calculate the paraxial image location for past flat lens implementations - operating in both the near- and far-fields - to test whether the existence of a real paraxial image can be used as  a general criterion to predict flat lens behavior.  We then apply the prerequisite of a real paraxial image location as a metric to explore new flat lens configurations.

The paraxial image location is derived from PF phase profiles calculated using the transfer matrix method~\cite{Yeh1988,Kong2005}.  We compare predictions of the image plane location to full-wave electromagnetic simulations, which are performed using the finite-difference frequency-domain (FDFD) technique to solve Maxwell's equations over a two-dimensional grid with pixel size of at most 1 nm.  For each simulation, the flat lens structure will image a sub-diffractive test object defined by two, $\lo/10$-wide openings spaced by a sub-wavelength distance $\lo/2.5$ in an opaque chromium masking layer (below the diffraction limit).  The simulations employ a sub-diffractive object because it provides additional information on the super-resolution capabilities of each flat lens, a topic of intense contemporary interest.  The object is illuminated at normal incidence by a plane wave of either TM or TE polarization propagating along the $+z$ direction.  The results of the simulations are presented in terms of the distribution of magnetic or electric energy density, depending on the use of TM or TE polarization, respectively.  Transfer matrix method calculations and FDFD simulations use permittivity values of silver and gold from~\cite{Johnson1972} and chromium from~\cite{CRCHandbook}.  Calculations corresponding to past published flat lens implementations use permittivity values reported with the results, where possible.

\section{Comparison with Past Flat Lens Results}
\begin{figure}[h]
  \begin{center}
    \includegraphics[scale = 0.65]{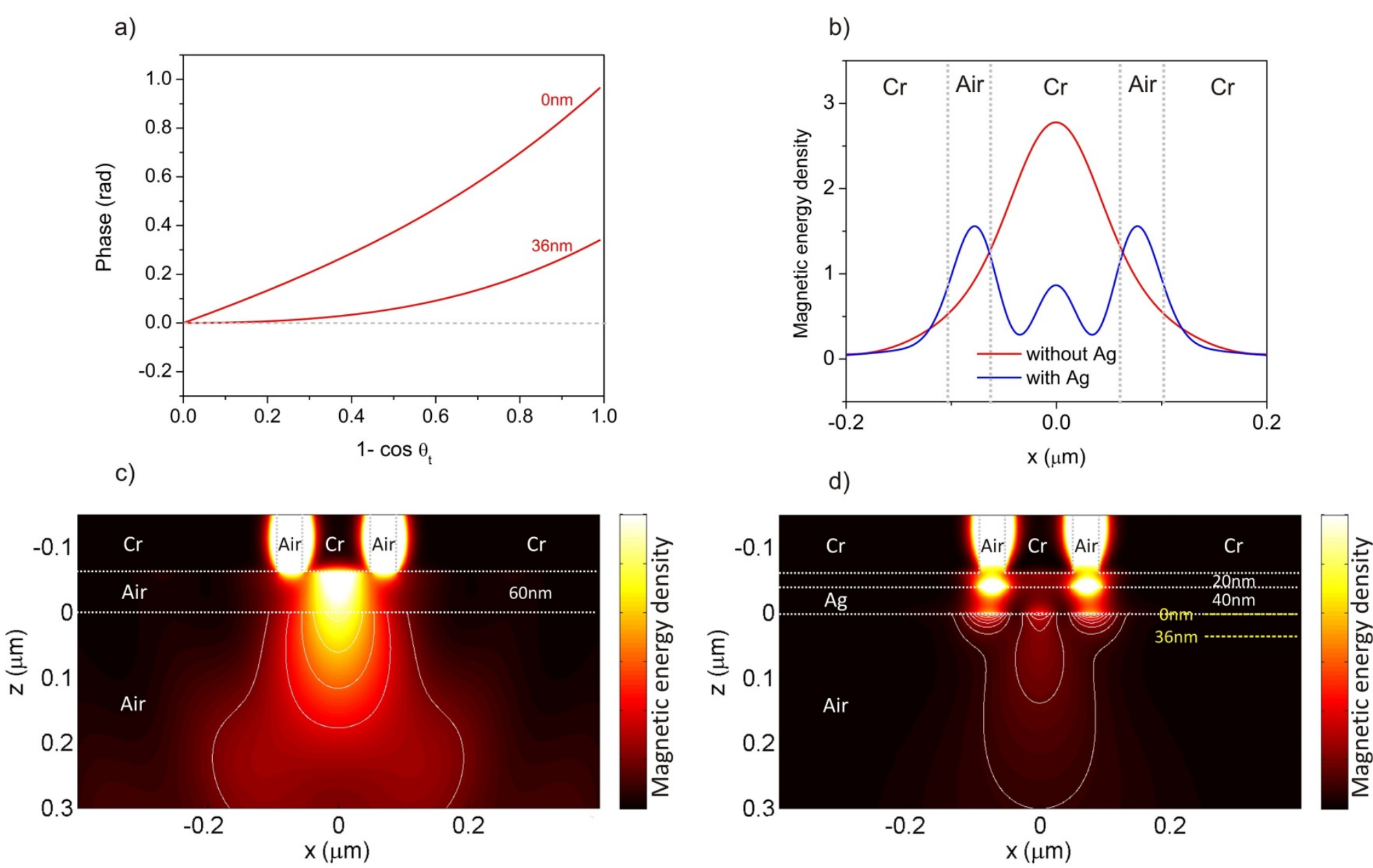}
    \caption[]{ (a) PF phase for Pendry's silver slab lens consisting of a 40-nm-thick Ag layer with a permittivity of $\ce = -1.0 + 0.4 i$ at a wavelength of $\lo = 356.3\nm$, where the object is 20 nm away from the entrance of the slab.  The phase and amplitude have been calculated at the exit of the slab ($z$ = 0 nm) and the predicted paraxial image location ($z$ = 36 nm).  (b) FDFD-simulated profile of the magnetic energy density at the image plane $z$ = 36 nm for the cases where the sub-diffractive object is imaged without (blue) and with (red) the silver slab lens. Simulated time-averaged magnetic energy density distributions of the illuminated object are shown (c) without and (d) with the 40-nm-thick Ag layer. The yellow dashed lines in (d) show the positions where the PF phase profiles have been calculated in (a).  \label{Fig3}}
  \end{center}
\end{figure}

\subsection{Pendry's Silver Slab Lens}
We begin by calculating the paraxial image location of the simplest configuration of a thin silver layer in air. It was first shown by Pendry that this type of flat lens could be a substitute for a Veselago lens composed of a $n=-1$ slab and provide a feasible means to image beyond the diffraction limit by evanescent wave amplification~\cite{Pendry2000}.  Although small-angle phase cannot directly predict super-resolution behavior, it can be used to predict the existence and location of its paraxial image, whose relationship with the super-resolved image has yet to be fully studied.  We consider the specific setup proposed by Pendry of an object in air spaced 20 nm from a 40-nm-thick silver layer, which is assumed to have a complex permittivity $\ce = -1.0 + 0.4 i$ at a wavelength of $\lo = 356.3\nm$.  For this case, the flat lens criterion predicts that the layer generates a real paraxial image located at $s = 36\nm$, comparable to the nominal image location of 20 nm used in~\cite{Pendry2000}.  Figure~\ref{Fig3}(a) shows the phase versus $q_t$ at the exit of the layer and the paraxial image location.  The positive slope of the phase at the exit of the layer indicates the existence of a real paraxial image, whose location corresponds to the plane $s = 36\nm$ where the phase profile at small angles is flat. Comparative simulations of the image of the sub-diffractive object - with and without the silver slab lens - reveals the super-resolving capabilities of the slab at the paraxial image location [Fig.~\ref{Fig3}(b)].  The spatial energy density distributions for the two cases, shown in Figs.~\ref{Fig3}(c) and \ref{Fig3}(d), further highlight super-resolution imaging conferred by the thin silver layer.  Although decay of the energy density near the exit face makes it difficult to definitively locate the image plane location, the predicted paraxial image location of $s = 36\nm$ is nonetheless consistent with the well-defined image of the sub-diffractive object localized near the slab exit.

\subsection{Near-Field Imaging with Silver Layers}
\begin{figure}[h]
  \begin{center}
    \includegraphics[scale = 1.2]{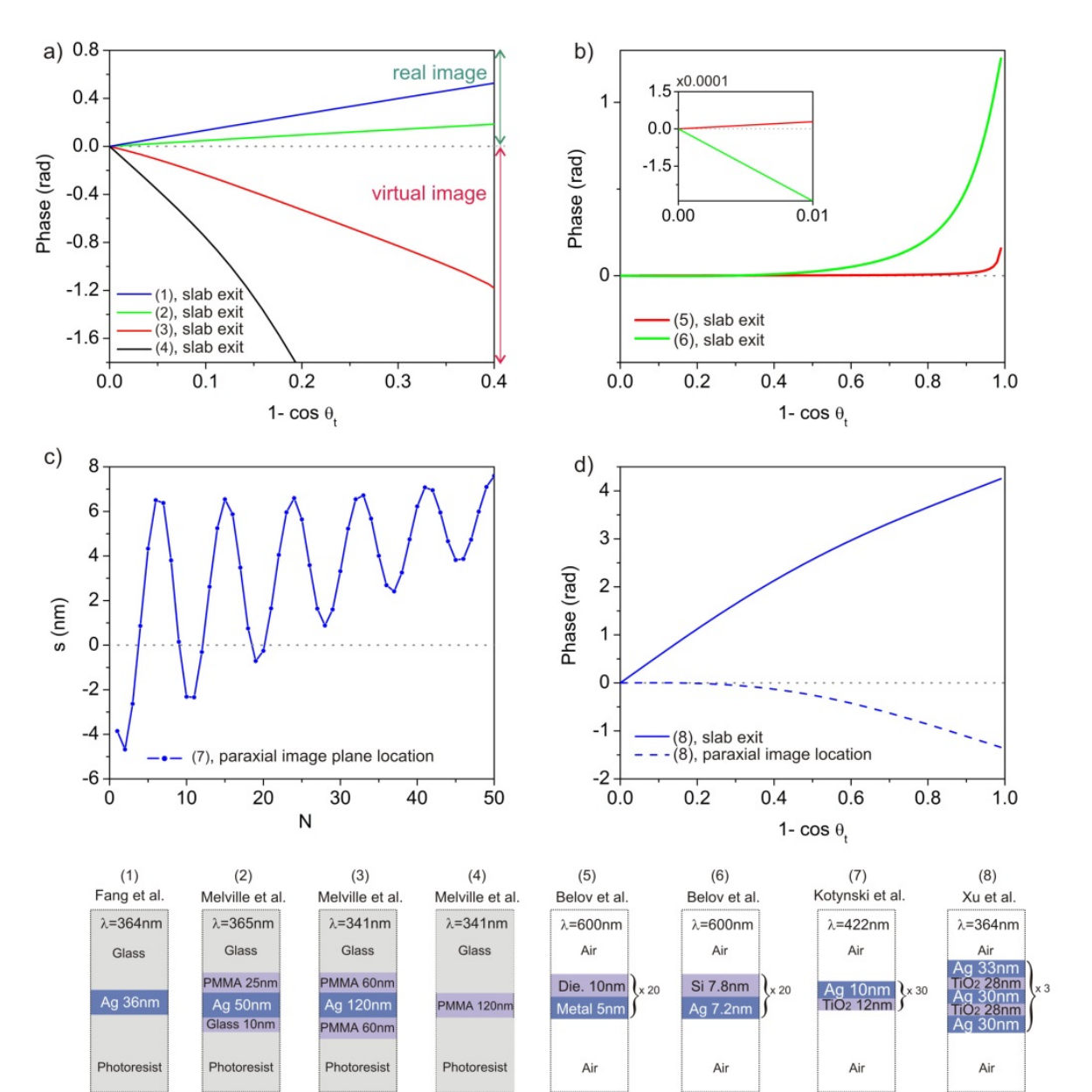}
    \caption[]{Flat lens criterion applied to past implementations.  (a) PF phase at the exit surface of lenses based on the 36-nm-thick silver layer studied in Fang et al. \cite{Zhang2003}, the 50-nm-thick silver layer studied in Melville et al. \cite{Melville2005}, and the 120-nm-thick silver layer studied by Melville et al. in \cite{Melville2004,Melville2004B}, along with the control used in \cite{Melville2004B} of a 120-nm-thick PMMA layer.  (b) PF phase at the exit surface of lenses based on metal-dielectric multi-layers studied by Belov et al. \cite{Belov2006}.  The inset in (b) shows a magnified view of the data near normal incidence.  (c) Paraxial image location as a function of unit cell repetition for the periodic metal-dielectric layered system studied Kotynski et al. \cite{Kotynski2011APA}. (d) PF phase at the exit surface for a flat lens based on a metal-dielectric layered system studied by Xu et al. \cite{Lezec2013}.  \label{Fig4} }
  \end{center}
\end{figure}

A correlation between the real paraxial image and the super-resolution image created by near-field flat lenses is further supported by analyzing past experiments showing UV images imprinted by a silver layer in photoresist.  Earlier experiments using silver layers of thicknesses up to 120 nm achieved images that were not sub-diffractive~\cite{Melville2004,Durant2005}, but noticeably sharper than those produced by a control sample consisting of a 120-nm-thick PMMA layer~\cite{Melville2004B}.  Analyzing their paraxial image locations provides a plausible explanation for these observations.  As shown in Fig.~\ref{Fig4}(a), the 120-nm-thick silver layer at a wavelength of $\lo=341\nm$ produces a negative-sloped phase profile, indicating a virtual paraxial image located at $s = -40 \nm$ outside of the photoresist region. Although the 120-nm-thick Ag layer does not produce a real paraxial image, it is nonetheless an improvement over the control of a 120-nm-thick PMMA layer, which also has a negative-sloped phase profile corresponding to a virtual paraxial image located at $s= - 150 \nm$.  In similar experiments, it was shown that thinner silver layers of thicknesses down to 35 nm could image features that were beyond the diffraction limit~\cite{Zhang2003,Zhang2005Science, Zhang2005NJP,Melville2005}.  As shown in Fig.~\ref{Fig4}(a), the 36-nm-thick and 50-nm-thick silver layers used in \cite{Zhang2003} and \cite{Melville2005} at UV wavelengths both have positive-sloped phase profiles and that respectively indicate real paraxial images located at $s=10 \nm$ and $s = 22 \nm$ in the photoresist region.  This comparative analysis suggests that super-resolution imaging requires a real paraxial image within the image-capturing region, although further work is needed to fully understand and validate this relationship.

\subsection{Anisotropic Metamaterial Lenses}
Flat lens implementations using repeated metal-dielectric bi-layers have been presented on the premise that they are anisotropic metamaterials, mimicking the flat lens conditions laid out in past proposals~\cite{Smith2004,Lu2005}.  If the composition of the metal-dielectric bi-layer satisfies the relations $\eM/\eD = -\dM/\dD$ and $\eM + \eD = 1$, where $\eM$ ($\eD$) and $\dM$ ($\dD$) are the respective permittivity and thickness values of the metal (dielectric) layer, the bi-layer can be described by a highly anisotropic permittivity tensor characterized by a flat wave vector diagram~\cite{Belov2006,Elson2006,Benedicto2012}.  This flat wave vector relation enables a form of imaging characterized by direct image projection from the front to the back of the slab.

We next test the consistency of the paraxial image location and the reported imaging properties of three different multi-layered systems designed as anisotropic metamaterial lenses in~\cite{Belov2006,Kotynski2011APA}.  PF phase [Fig. \ref{Fig4}(b)] calculated for the two lossless 20-unit-cell systems studied in \cite{Belov2006} at the wavelength $\lo = 600\nm$, one using unit cell parameters $\eM=-1$, $\eD=2$, and $\dM/\dD = 1/2$ and the other $\eM=-14$, $\eD=15$, and $\dM/\dD = 14/15$, yield respective paraxial image locations of $s = 0.2\nm$ and $s = -3\nm$.  These small paraxial image locations are consistent with reported observations of image projection on the back of the slabs.  In a different study, detailed simulation data provided for the lossy multi-layered system presented in \cite{Kotynski2011APA} at the wavelength $\lo = 422\nm$, using unit cell parameters $\eM=-5.637 + 0.214 i$, $\eD=2.6^2$ and $\dM/\dD = 10/12$, revealed an oscillatory dependence of the spot size at a fixed plane (near the output face) consisting of 6 evenly spaced local minima as the number of repetitions $N$ increases from 1 to 50.  The paraxial image location calculated by PF phase for this system [Fig. \ref{Fig4}(c)] exhibits an identical oscillatory dependence on $N$, suggesting that the spot size variation observed in \cite{Kotynski2011APA} can at least be partially attributed to image focusing and de-focusing about the output face.

\subsection{Negative-Index Metamaterial Lens}
Recently, a flat lens consisting of metal-dielectric multi-layers has been presented as a metamaterial with an isotropic negative index~\cite{Lezec2013}, mimicking the original flat lens proposal by Veselago~\cite{Veselago1968}.  PF phase [Fig. \ref{Fig4}(d)] calculated for this system at the wavelength $\lo = 364\nm$ confirms that it acts as a flat lens and predicts a paraxial image location of $s=370 \nm$, agreeing well with the experimentally measured image location of $360\nm$~\cite{Lezec2013}. Extrapolation of the phase profile to the paraxial image location reveals wavefront aberrations lower than $\lo/4$ up to about unity numerical aperture, an indication of high imaging quality consistent with the near-diffraction-limited images observed for this system.

\section{Validation of Flat Lens Criterion by Full-Wave Simulations}
\begin{figure}[h]
  \begin{center}
    \includegraphics[scale = 0.90]{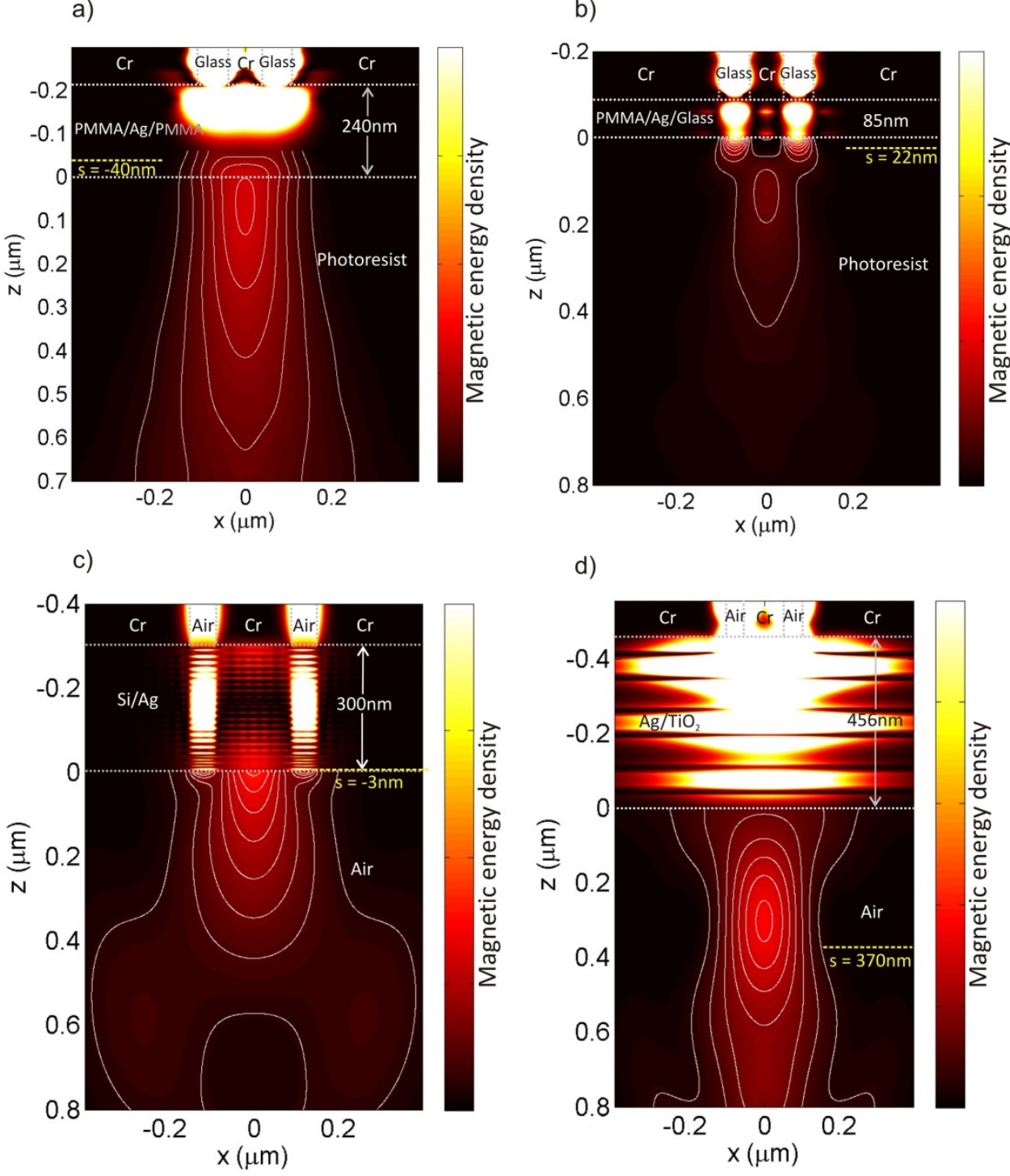}
    \caption[]{ Comparison of paraxial image locations predicted by PF phase and numerical simulations.  FDFD-calculated time-averaged energy density distributions for flat lenses consisting of (a) a 120-nm-thick silver layer studied in~\cite{Melville2004,Melville2004B}, (a) a 50-nm-thick silver layer studied in~\cite{Melville2005}, (c) metal-dielectric multi-layers studied in~\cite{Belov2006}, and (d)  metal-dielectric multi-layers studied in~\cite{Lezec2013}.  In all cases, we use a sub-diffractive object consisting of two, $\lo/10$-wide openings spaced $\lo/2.5$ apart in an opaque mask that is illuminated by a TM-polarized plane wave. The yellow dashed line in each panel shows the corresponding paraxial image location calculated from the slope of the output phase. \label{Fig5} }
  \end{center}
\end{figure}

Full-wave electromagnetic simulations of past flat lens configurations reveal consistent agreement between the paraxial image location and the simulated image location across near-field and far-field regimes.  We consider four different lenses, all for TM-polarization, that have been examined in the previous section: 1) the 120-nm-thick silver layer studied in~\cite{Melville2004,Melville2004B}, 2) the 50-nm-thick silver layer studied in~\cite{Melville2005}, 3) the anisotropic metamaterial lens based on repeated metal-dielectric layers studied in~\cite{Belov2006}, and 4) the negative-index metamaterial lens also based on repeated metal-dielectric layers studied in~\cite{Lezec2013}.  Simulations of the systems are shown in Fig.~\ref{Fig5}, with the paraxial image locations depicted by dashed yellow lines.  The lens based on the 120-nm-thick silver layer yields a divergent energy density distribution in the image region with its narrowest plane located close to the virtual paraxial image location $s = -40\nm$ [Fig.~\ref{Fig5}(a)].  On the other hand, the lenses based on the 50-nm-thick silver layer and anisotropic metamaterial both exhibit sub-diffractive images of the object located near their output faces, at positions closely matching the real paraxial image locations predicted by PF phase [Figs.~\ref{Fig5}(b) and \ref{Fig5}(c)]. This agreement is surprising because the paraxial image location is determined without considering large-angle components believed to be the main culprit of super-resolution imaging.  Finally, the negative-index metamaterial lens generates a diffraction-limited image of the object well-separated from the output face at a position that again agrees well with the real paraxial image location [Fig.~\ref{Fig5}(d)].  Unlike the other lenses, this lens is diffraction limited because the energy density distribution in the image region is incapable of resolving the two point-like openings of the object.

\section{Flat Lens Condition for Thin Layers}

We develop compact analytical expressions from Eq.~(\ref{paraxialimage}) to delineate when a single homogeneous layer can generate a real paraxial image, whose location has been shown in the previous section to be strikingly consistent with even near-field super-resolved images.  To begin, we take the most generic case of a homogeneous, non-magnetic, isotropic layer of thickness $d$ where the bounding half spaces are free space.  The layer is characterized by a permittivity $\ce = \e' + i \e''$ and a permeability $\mu = 1$.  The response of the layer to a point source placed on one face of the layer can be analyzed in terms of a plane-wave spectrum parameterized by the wave vector in the free-space image region given by $\vko = \kox\xh + \koz\zh$.  Each plane-wave component exiting the layer is modified by a transmission coefficient
\begin{equation}
\ct = \frac{4 \cp }{(1+\cp)^2 e^{-i\ckz d} - (1-\cp)^2 e^{i\ckz d}} \quad,
\end{equation}
where $\ckz = \ko \sqrt{\cn^2-\sin\theta^2}$ is the propagation constant in the layer, $\theta$ is the angle of incidence, $\ko$ is the magnitude of the free-space wave vector, $\cn = \sqrt{\ce}$ is the layer refractive index, and $\cp$ is a parameter defined as $\ckz/(\ce \koz)$ for TM polarization and $\ckz/\koz$ for TE polarization.  The transmission coefficient accounts for infinite multiple reflections and propagation loss within the slab, and its phase corresponds to the PF phase $\Phiz(z=0)$ from which the paraxial image can be estimated.

\begin{figure}[h]
  \begin{center}
    \includegraphics[scale = 0.7]{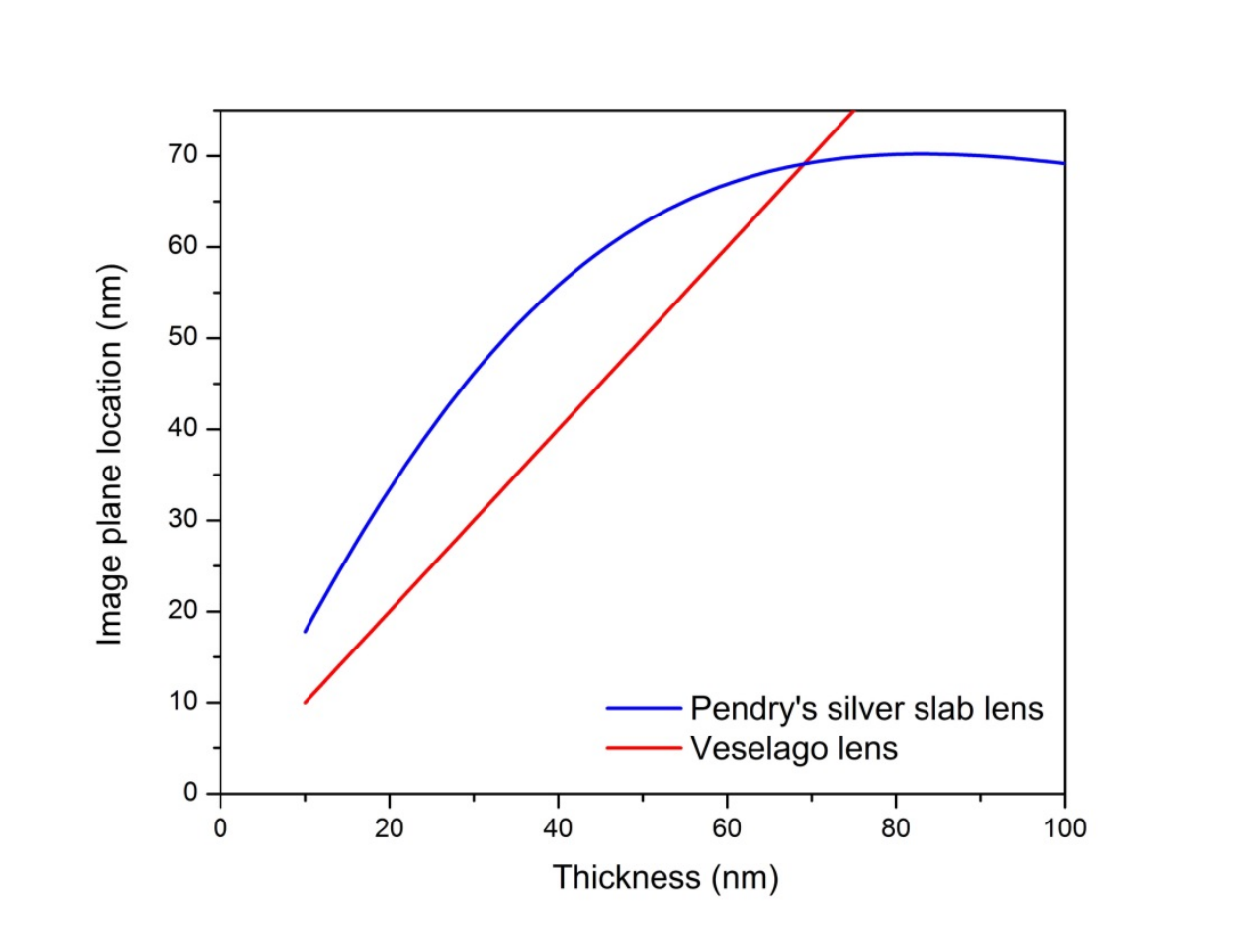}
    \caption[]{ Paraxial image location versus thickness for an illuminated object located at the entrance of the ideal Veselago lens (red) and Pendry's silver slab lens (blue) when illuminated at the wavelength of $\lo = 356.3 \nm$.  For the ideal Veselago lens, the image location is equivalent to the slab thickness, $s = d$. \label{Fig6}}
  \end{center}
\end{figure}

We first re-visit Pendry's predictions of a silver slab lens by deriving a flat lens criterion for a lossy layer under the thin-film and paraxial approximations for TM polarization.  To first order of $\ko d$, this yields
\begin{equation} \label{FLTM1}
\frac{\partial \PhizTM(z=0)}{\partial q_t}\Big\vert_{q_t=0} \doteq -\ko d \left( \frac{1}{2}(\e'-1) + \frac{\e'}{\|\ce\|^2}\right) \quad,
\end{equation}
which must be positive to realize real image formation.  This occurs if $\e'<|\ce|^2/(|\ce|^2 + 2)$, a condition that can be satisfied by metals.  For Pendry's silver slab lens with $\ce = -1 + 0.4 i$~\cite{Pendry2000}, Eq.~(\ref{FLTM1}) predicts a paraxial image location $s = 1.86 d$ that is nearly double the expected image location for a Veselago lens of equivalent thickness. Figure~\ref{Fig6} maps the thickness dependence of the paraxial image location of an object placed directly on the entrance of an ideal Veselago lens and Pendry's silver slab lens.  Although the two are within the same order of magnitude, discrepancies highlight the limitation of equating their imaging properties even under electrostatic conditions.

\label{losslesslayer}

We next develop an expression to qualitatively explain why thinner silver layers are better able to fulfill the flat lens criterion.  Ignoring losses in silver (a rough approximation of its response at UV frequencies), the flat lens criterion under the thin-film and paraxial approximations for TM polarization is given up to third order of $\ko d$ by
\begin{equation}\label{slopeTMnoloss}
\frac{\partial \PhizTM(z=0)}{\partial q_t}\Big\vert_{q_t=0}  \doteq -\ko d \frac{\e'^2 -\e' +2}{2\e'} + (\ko d)^3 \frac{(\e'-1)^2(3\e'^2+5\e'+6)}{24\e'},
\end{equation}
which must be positive to realize real image formation.  Consideration of just the first term in Eq.~(\ref{slopeTMnoloss}) would lead to the conclusion that a flat lens simply requires $\e'<0$.  However, the higher order term in Eq.~(\ref{slopeTMnoloss}) has a sign opposite to that of the first term, indicating that increases in the thickness of a negative permittivity layer generally counteract phase conditions necessary to make a flat lens.  Given the connection we have shown between a real paraxial image and a super-resolution image, this could explain why an early attempt to make a flat lens using a thicker silver layer did not achieve super-resolution~\cite{Melville2004,Durant2005}, whereas later attempts using thinner silver layers did~\cite{Zhang2005Science,Zhang2005NJP,Melville2005}.

\section{Flat Lens for TE Polarization}

The generality of the flat lens condition enables exploration of the possibility of flat lenses for TE polarization.  Taking the simplest case of a lossless dielectric layer, the flat lens criterion  under the thin-film and paraxial approximations for TE polarization is given by
\begin{equation}\label{slopeTEnoloss}
\frac{\partial \PhizTE(z=0)}{\partial q_t}\Big\vert_{q_t=0} \doteq -\ko d \frac{3 - \e'}{2} - (\ko d)^3 \frac{(\e'-1)^2 (3\e'-1)}{24}.
\end{equation}
Considering only the first term of Eq.~(\ref{slopeTEnoloss}), a dielectric layer can act as a flat lens if $\e' > 3$, a condition satisfied by many types of glasses and semiconductors.  The higher order term in Eq.~(\ref{slopeTEnoloss}) has a sign that is opposite to that of the first order term.  This suggests that although a flat lens for TE polarization is possible with a dielectric layer, the thickness of the layer must be sufficiently small to limit deleterious phase contributions from the third-order term.

\begin{figure}[h]
  \begin{center}
    \includegraphics[scale = 1.03]{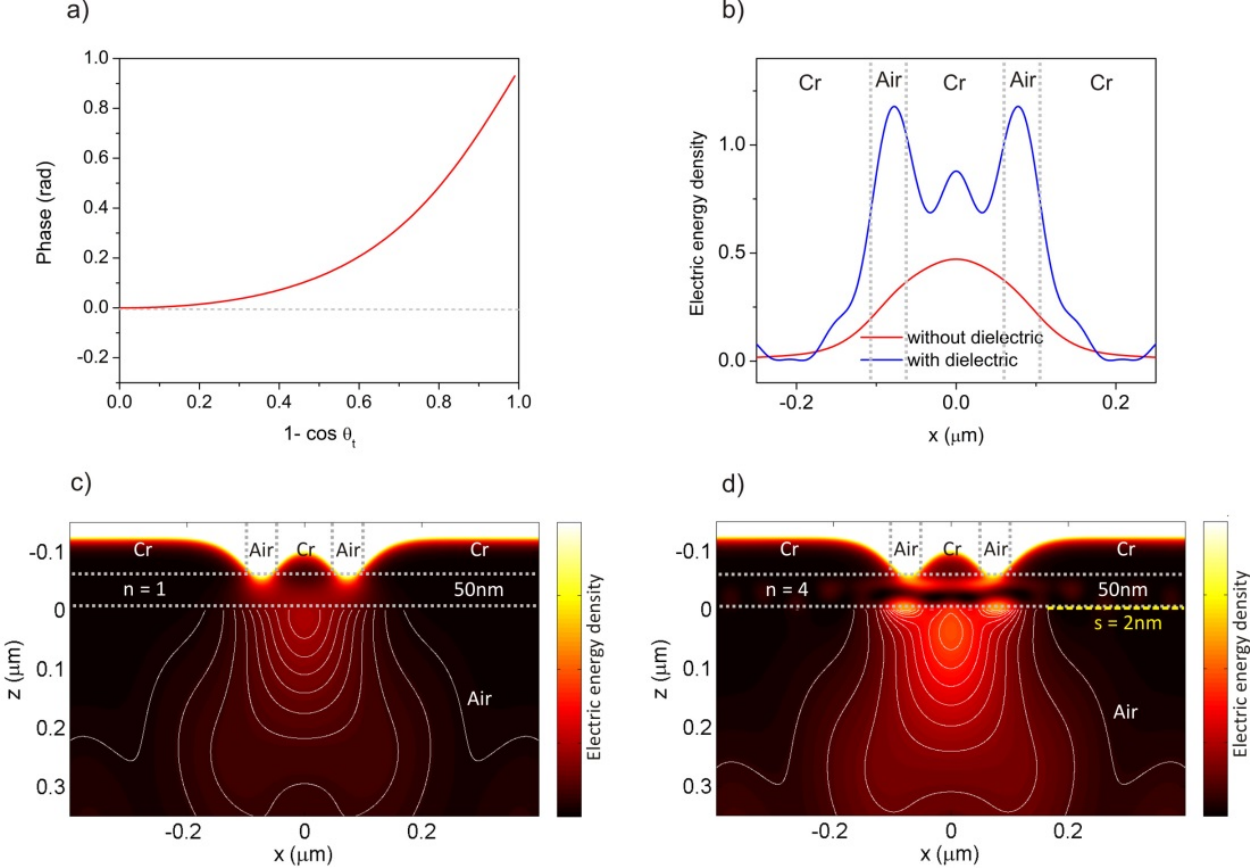}
    \caption[]{ Flat lens for TE polarization based on a 50-$\nm$-thick lossless dielectric ($n=4$) layer immersed in air and illuminated at a wavelength $\lo=365\nm$.  (a) PF phase at the paraxial image location $s = 2 \nm$.  (b) FDFD-simulated profile of the electric energy density at the paraxial image location for the cases where the object is imaged without (blue) and with (red) the dielectric slab. Simulated time-averaged electric energy density distribution of the illuminated object are shown (c) without and (d) with the 50-nm-thick dielectric layer.  The yellow dashed line in each panel shows the paraxial image location calculated by PF phase.  \label{Fig7}}
  \end{center}
\end{figure}

To further explore the possibility of a flat lens for TE polarization based on a dielectric layer, we study the configuration of a $50$-$\nm$-thick high-index dielectric ($n=4$) layer immersed in air at a wavelength of $365\nm$.  PF phase analysis predicts a real paraxial image location $s=2\nm$.  The phase and amplitude at this location are shown in Fig.~\ref{Fig7}(a).  We perform two FDFD simulations of a sub-diffractive object imaged with and without the dielectric layer.  As shown in Fig.~\ref{Fig7}(b), the addition of the dielectric layer yields an image of the sub-diffractive object at the paraxial image location.  This appears to be evidence of super-resolution imaging for TE polarization, but it should be cautioned that the effective wavelength in the dielectric layer has reduced by a factor of 4 and that, from the perspective of the layer, the objects are no longer spaced at a sub-diffractive distance.  Simulations [Figs.~\ref{Fig7}(c) and \ref{Fig7}(d)] further highlight the ability of the dielectric layer to make a near-field image of the two point-like openings over its extent.  The location of this image is once again consistent with the real paraxial image location predicted to be just a few nanometers from the output face.

\section{Flat Lens Condition for Multi-Layers}

We next develop an expression for the flat lens condition for a multi-layered system, showing that it matches a condition used to make anisotropic metamaterial lenses for the case of a bi-layer unit cell.  Based on Eq.~(\ref{FLTM1}), the small-angle phase behavior of a multi-layered system for TM polarization is approximately
\begin{equation}\label{FLTM2}
\frac{\partial \PhizTM(z=0)}{\partial q_t}\Big\vert_{q_t=0} \doteq -\ko \sum_{i} \di \left( \frac{1}{2}(\ei'-1) + \frac{\ei'}{\|\cei\|^2} \right) \quad,
\end{equation}
where $\di$ is the thickness and $\cei = \ei' + \ei''$ is the permittivity of the $i^{\rm th}$ layer.  From Eq.~(\ref{FLTM2}), it is possible to design a metal-dielectric bi-layer unit cell capable of TM-polarized image projection across its extent by imposing a plane curvature to the wavefront transmitted through a unit cell.  For a metal-dielectric bi-layer unit cell with parameters $\dD$, $\eD$, $\dM$, and $\ceM=\eM' + \eM''$ (where the metal permittivity is now allowed to take on complex values), the required dielectric permittivity to achieve image projection across the unit cell is given by
\begin{equation} \label{anisocond}
\eD = \frac{1}{2} - \frac{\dM \cgM }{\dD} \pm \sqrt{ \left( \frac{1}{2} - \frac{\dM \cgM }{\dD}  \right)^2 -2  } \quad,
\end{equation}
where $\cgM = (\eM' - 1)/2 + \eM'/|\ceM|^2$.  In the case of a real, large metal permittivity $\eM' >> 1$ and comparable layer thicknesses $\dM = \dD$, the condition given by Eq.~(\ref{anisocond}) simplifies to $\eM/\eD = -\dM/\dD$ and $\eM + \eD = 1$, exactly matching the design constraints imposed by effective medium theory to achieve a metamaterial with an anisotropic permittivity tensor~\cite{Belov2006}.

\section{Broadband Flat Lens Designed by Small-Angle Phase}

\begin{figure}[h]
  \begin{center}
    \includegraphics[scale = 1.0]{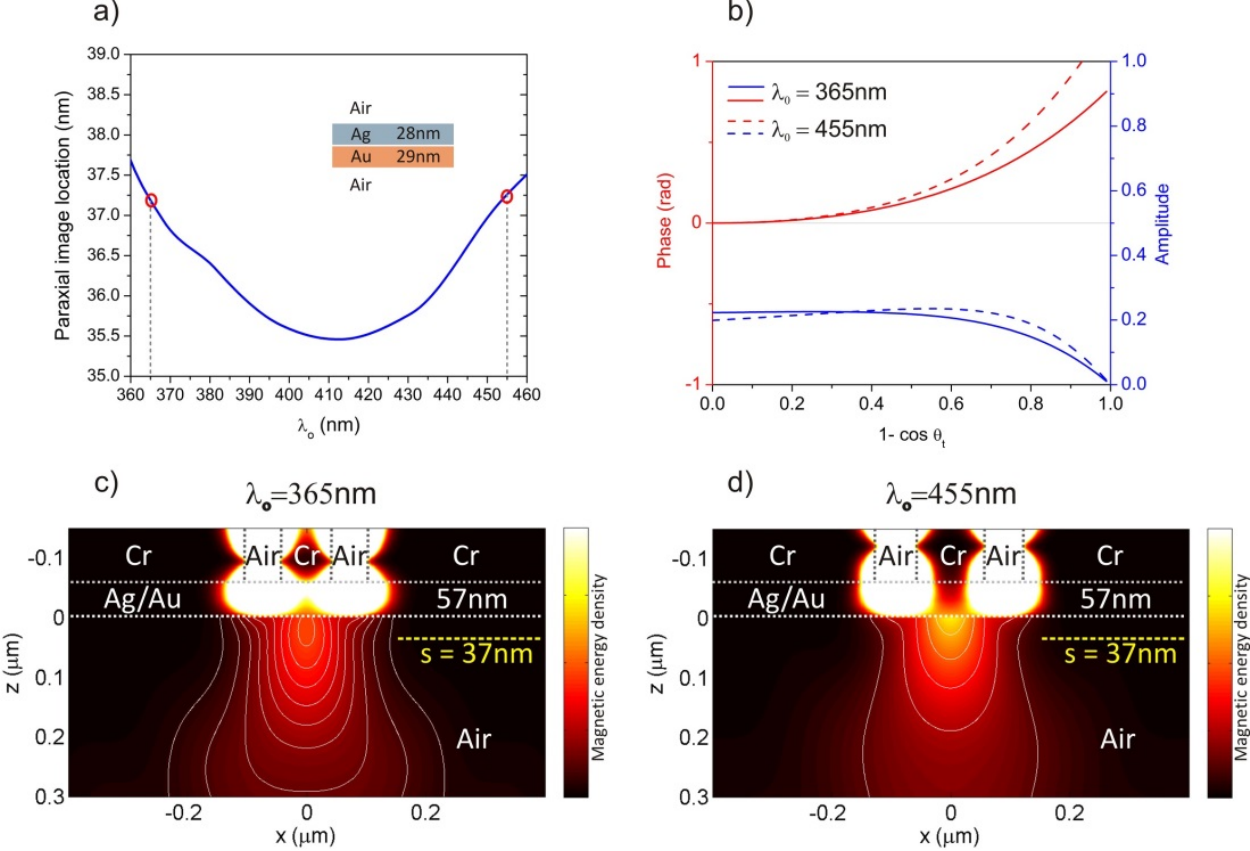}
    \caption[]{ Engineering a broadband flat lens.  (a) Paraxial image location over the ultraviolet-blue spectrum for a bi-layer flat lens consisting of a 28-$\nm$-thick silver layer and a 29-$\nm$-thick gold layer immersed in air.  (b) PF phase (red) and amplitude (blue) for the bi-layer flat lens at the wavelength of $\lo = 365\nm$ (solid lines) and $\lo = 455\nm$ (dashed lines). Time-averaged energy density distributions for the bi-layer system under plane-wave illumination at (c) $\lo = 365\nm$ and (d) $\lo = 455\nm$.  The yellow dashed line in each panel shows the paraxial image location calculated by PF phase. \label{Fig8}}
  \end{center}
\end{figure}

Using the paraxial image location in Eq.~(\ref{paraxialimage}) as a merit function, we implement an optimization routine to systematically design a bi-layer system with a consistent image plane location for TM polarization over a large portion of the ultraviolet-visible spectrum.  One result of this design process is a bi-layer consisting of a 28-$\nm$-thick silver layer and a 29-$\nm$-thick gold layer (this is not the only solution, as many configurations can possess the same image location).  The paraxial image location of the bi-layer remains relatively constant over a large spectral range, varying between $35\nm$ and $37\nm$ over the range $365\nm<\lo<455\nm$ [Fig.~\ref{Fig8} (a)].  Figure~\ref{Fig8}(b) shows the PF amplitude and phase at the paraxial image location $s=37\nm$ at the wavelengths of $\lo=365\nm$ and $\lo=455\nm$.  Full-wave electromagnetic simulations [Fig.~\ref{Fig8}(c) and \ref{Fig8}(d)] modeling the response of the bi-layer at the lower and upper bounds of this wavelength range reveal similar energy density distributions that have tapers located near the predicted paraxial image locations of $37\nm$.  Note that unlike other near-field flat lenses, the bimetallic lens is not capable of resolving the sub-diffractive object.  We attribute the loss of resolution to augmented metallic losses due to the addition of gold.

\section{Far-Field Immersion Flat Lens}
\begin{figure}[h]
  \begin{center}
    \includegraphics[scale = 1.0]{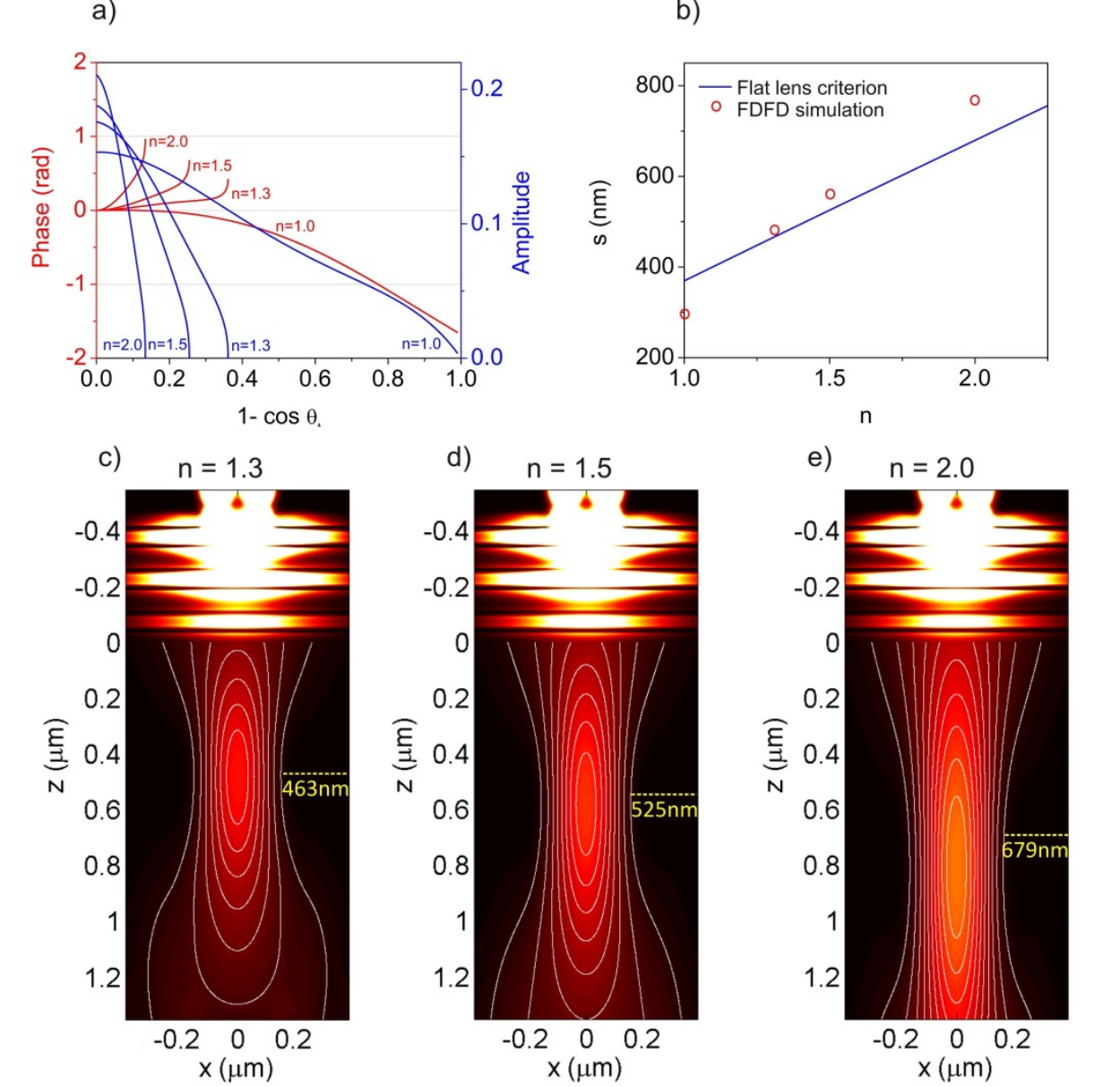}
    \caption[]{ Enhancing the image plane location of the multi-layered flat lens system previously studied in~\cite{Lezec2013} by immersion of the image region in a dielectric.  (a) PF phase (red) and amplitude (blue) at the paraxial image location for the cases where the dielectric medium has refractive index $n$ = 1.0, 1.3, 1.5, and 2.0. (b) Paraxial image location versus the refractive index of the dielectric medium predicted by PF phase (blue line) and FDFD simulations (red circles).   (c), (d), and (e) show FDFD-calculated magnetic energy density distributions of the immersed flat lens system for $n$ = 1.3, 1.5, and 2.0, respectively.  The yellow dashed lines in panels (c)-(e) show the paraxial image location calculated by PF phase.  \label{Fig9}}
  \end{center}
\end{figure}

A simple method to boost the paraxial image location is to immerse the image region in a dielectric. We apply this technique to further extend the far-field image location of the multi-layered system studied in~\cite{Lezec2013}, which already sets the current benchmark for the largest image-to-lens separation.  As shown in Fig.~\ref{Fig9}(b), the paraxial image location for this system linearly ramps up as a function of the refractive index of the immersing dielectric.  This effect can be visualized through simulations [Figs.~\ref{Fig9}(c)--\ref{Fig9}(e)], which show an increasing separation between the energy density maxima in the image region and the exit of the lens as the refractive index of the image region increases.  Good agreement is found between the position of the maxima obtained from simulations and the paraxial image location predicted by PF phase [Fig.~\ref{Fig9}(b)].  Although enhancements in the image plane location come at the cost of increased abberations - as shown by the PF amplitude and phase at the paraxial image location for several refractive index values [Fig.~\ref{Fig9}(a)] - dielectric immersion nonetheless provides an elegant method to boost the image location of flat lenses and increase its working distance.  This is important if flat lenses are to be eventually used to image three-dimensional objects.

\section{Discussion}

The flat lens criterion proposed in this work offers a single metric for predicting real image formation by stacks of layers composed of homogeneous and isotropic media and can potentially be used for rapid flat lens design over large parameter spaces.  However, the metric does not supplant existing methods of flat lens analysis and its limitations should be noted.  One limitation is that the flat lens criterion has been derived without considering interactions between the object and lens, which have been shown to be significant when the object-to-lens separation is small~\cite{Moore2009}.  Incorporating these interactions would likely lead to more accurate predictions, but sacrifices generality because each prediction becomes object-dependent.  Another limitation is that the flat lens criterion is based on small-angle phase alone and cannot provide information on the resolution, contrast, or fidelity of the image.  Alternative criteria can be established to examine large-angle plane-wave components with NA near unity or evanescent plane-wave components with NA greater than one, which are worthy of future studies.  The flat lens criterion is also limited to configurations in which the coherence length of the light source is much larger than the thickness of the layered stack.  The influence of coherence on imaging quality should be further investigated to assess the compatibility of flat lenses with different types of light sources.  Flat-lens imaging relies on the interference of multiply reflected waves in the stack, which is best realized using coherent laser light and utterly impossible using totally incoherent light.  Given sufficiently thin stacks, it should be possible to image using partially coherent light from narrow-band light-emitting diodes, a configuration that is low-cost and amenable to fluorescence imaging.  Finally, a surprising and puzzling finding of this work is that the small-angle flat lens criterion (NA near zero) can accurately predict the location of images produced from super-resolving flat lens systems (NA greater than unity).  This correlation has yet to be fully understood and suggests a need for a complete aberration theory for flat lenses that accommodates both propagating and evanescent components.

\section{Conclusion}
We have developed a flat lens criterion based on small-angle phase that finds general consistency with implementations to date and, as a result, reveals basic small-angle phase behavior shared by many flat lenses.  Analytical expressions for the flat lens criterion under various conditions reveal the new possibility of flat lenses for TE polarization, with a case examined here that is capable of super-resolution imaging.  Systematic flat lens design is now possible, which we have demonstrated through the design of a broadband flat lens over the UV-visible spectrum and an immersion lens with an adjustable far-field paraxial image location.  Future work should focus on engineering new flat lenses with desirable, but practically challenging, features such as operation over the entire visible spectrum or simultaneously imaging with TM and TE polarizations.

\section*{Acknowledgments}
It is a pleasure to acknowledge helpful comments from Andr\'{e} Phillion and L{\"o}ic Markley from The University of British Columbia, Amit Agrawal from Syracuse University, and Vladmir Aksyuk and Alex Liddle from the National Institute of Standards and Technology. This work was supported by the Natural Science and Engineering Council of Canada (NSERC) through Discovery Grant 366136.

\end{document}